\documentclass[twocolumn,preprintnumbers,amsmath,amssymb,floatfix,prd]{revtex4}
\usepackage{epsfig}
\usepackage{dcolumn}
\begin{document}
\title{Charged Hadron Fragmentation Functions at High Energy Colliders}
\author{Ignacio Borsa}
\email{ignacio.borsa@itp.uni-tuebingen.de} 
\author{Marco Stratmann}
\email{marco.stratmann@uni-tuebingen.de}
\affiliation{Institute for Theoretical Physics, University of T\"ubingen, Auf der 
Morgenstelle 14, 72076 T\"ubingen, Germany}
\author{Daniel de Florian}
\email{deflo@unsam.edu.ar} 
\affiliation{International Center for Advanced Studies (ICAS) and IFICI, UNSAM, 
Campus Miguelete, 25 de Mayo y Francia (1650) Buenos Aires, Argentina}
\author{Rodolfo Sassot}
\email{sassot@df.uba.ar} 
\affiliation{Universidad de Buenos Aires, Facultad de Ciencias Exactas 
y Naturales, Departamento de F\'{\i}sica and IFIBA-CONICET, Ciudad Universitaria, (1428) Buenos Aires, Argentina}
\begin{abstract}
We update our extraction of parton-to-charged hadron fragmentation functions 
at next-to-leading order accuracy in QCD, focusing on the wealth of data collected at 
the Large Hadron Collider over the past decade.  
We obtain an accurate description of single-inclusive processes involving unidentified 
charged hadrons produced at different rapidities and transverse momenta in proton-proton collisions 
in a wide range of center-of-mass system energies between 0.9 and 13 TeV, 
along with measurements performed in proton-antiproton collisions at the Tevatron Collider in the past. 
NLO estimates of charged hadron production rates agree 
best with data when the theoretical factorization scales are selected similar to those 
optimized for identified pions, kaons, and protons in a recent global QCD analysis.
\end{abstract}
%
\maketitle

\section{Introduction}
The production of charged hadrons with large transverse 
momentum $p_T$ in hadronic collisions is an ubiquitous tool used to explore different 
aspects of the structure of matter, its constituents, and their interactions. 
Applications range from the use of hadroproduction processes to unveil nuclear 
structure and the behavior of quark and gluons at very high energies in 
proton-nucleus and nucleus-nucleus collisions \cite{Klasen:2023uqj}, to the determination of, say, the quark and gluon polarization in polarized proton-proton and electron-proton collisions \cite{Borsa:2020lsz}. 

One cornerstone of these studies is the perturbative Quantum Chromodynamics 
(QCD) framework for single-inclusive high-$p_T$ hadron production \cite{Collins:1989gx} 
and, more specifically, its description in terms of fragmentation functions (FFs), as they link the 
hard scattering of partons at short distances to the hadrons observed in the final-state \cite{Field:1976ve}. 
Scale-dependent FFs factorize the relevant nonperturbative information on the hadronization 
process at long distances from the perturbative partonic cross sections, and 
cancel consistently their final-state singularities \cite{Collins:1981uw}.

Fifteen years ago, the first next-to-leading order (NLO) global QCD analysis of 
unidentified charged hadron FFs was presented in Ref.~\cite{deFlorian:2007ekg}. It combined 
data on single-inclusive electron-positron annihilation (SIA) with those from semi-inclusive 
deep-inelastic scattering (SIDIS) and proton-proton collisions (PP) and demonstrated
the universality and factorization properties of FFs within the precision of the 
data at that time. Similar analyses were later published in \cite{Bertone:2018ecm,Mohamaditabar:2018ffo,Soleymaninia:2020bsq,Moffat:2021dji} 
incorporating newer data and different theoretical refinements and analyses strategies.

Over the last decade, various LHC experiments 
\cite{ALICE:2010mty,CMS:2011mry,CMS:2012aa,ALICE:2013txf,LHCb:2021abm}
have produced remarkably precise 
hadroproduction data at increasing center-of-mass system (c.m.s.) energies. 
One rather unexpected feature of these data is the sizable discrepancy with 
NLO predictions computed with FFs obtained previously from SIA, SIDIS, and PP data 
at lower c.m.s.\ energies. 
The tension aggravates to the point that it is not 
even possible to describe consistently data sets from the same LHC experiment if the c.m.s.\ energies 
increases by a few TeV. This feature likely suggests a serious limitation of the
NLO framework and has also been observed in analyses of pion, kaon, and proton hadroproduction data
\cite{dEnterria:2013sgr}.

Nevertheless, in Ref.~\cite{Borsa:2021ran} it has been demonstrated that the 
NLO QCD framework can still provide a rather accurate global description of data with 
identified pions up to the highest c.m.s.\ energies available, and even
down to $p_T$-values of around 1 GeV, if one fully exploits the 
factorization scale ambiguity that is inherent to any fixed-order 
perturbative QCD estimate. In doing so, one can hope to mimic the yet 
unknown higher order QCD corrections, in particular, their energy dependence.
In the following, we show that this is also a viable path for analyzing
data for other identified or unidentified final-state hadrons, including
measurements in proton-antiproton collisions from the CDF experiment 
\cite{CDF:1988evs,CDF:2009cxa} and the even the older UA1 and UA2 experiments at the 
CERN Super Proton Synchroton \cite{UA1:1989bou,UA2:1984ida}. 
Interestingly, the correlation between the most appropriate choice of factorization scale 
and the c.m.s.\ energy is found to be the same as for pions in Ref.~\cite{Borsa:2021ran}.
In this way we are able to provide a new set of FFs for unidentified charged hadrons $h^{\pm}$ that is
suitable for up-to-date phenomenological applications at high energy hadron colliders. 

We note, that the new set of FFs is also constrained by very precise 
SIDIS data for unidentified charged hadrons from COMPASS \cite{COMPASS:2016xvm} 
that were not available at the time when the analysis of Ref.~\cite{deFlorian:2007ekg} was presented.
These results complement the pioneering measurements by EMC \cite{EuropeanMuon:1991sne} that were employed previously. 
As is customary, the estimated residual uncertainties of the FFs are given in terms of a large set of replicas 
obtained by Monte Carlo sampling, which easily propagate to any observable, including the 
measurements used in the fit. 

\section{Setup of our Global Analysis}
Since the methodology we employ to extract the charged hadron FFs is identical to 
the one used for pions, we refer the reader to \cite{Borsa:2021ran} and references therein 
for further details. In the following, we focus  
on the main features of the new data sets and the results of our analysis. 

The aforementioned tension between the NLO estimates for hadroproduction cross sections
and experimental results with increasing c.m.s.\ energy $\sqrt{s}$, is best illustrated by the CMS 
\cite{CMS:2011mry,CMS:2012aa} and ALICE \cite{ALICE:2010mty,ALICE:2013txf} data which are both
included in our analysis. The data correspond to values of $\sqrt{s}$  
comprising $0.9,\,2.76$, and $7\,\mbox{TeV}$ and a $p_T$-range up to 
around $200$ GeV. An advantage of comparing data from the same collaboration at different 
$\sqrt{s}$ is to minimize potential discrepancies stemming from varying
criteria used to identify and select certain final-state hadrons experimentally due to 
the underlying detector capabilities.

\begin{figure}[t]
\vspace*{-0.3cm}
\epsfig{figure=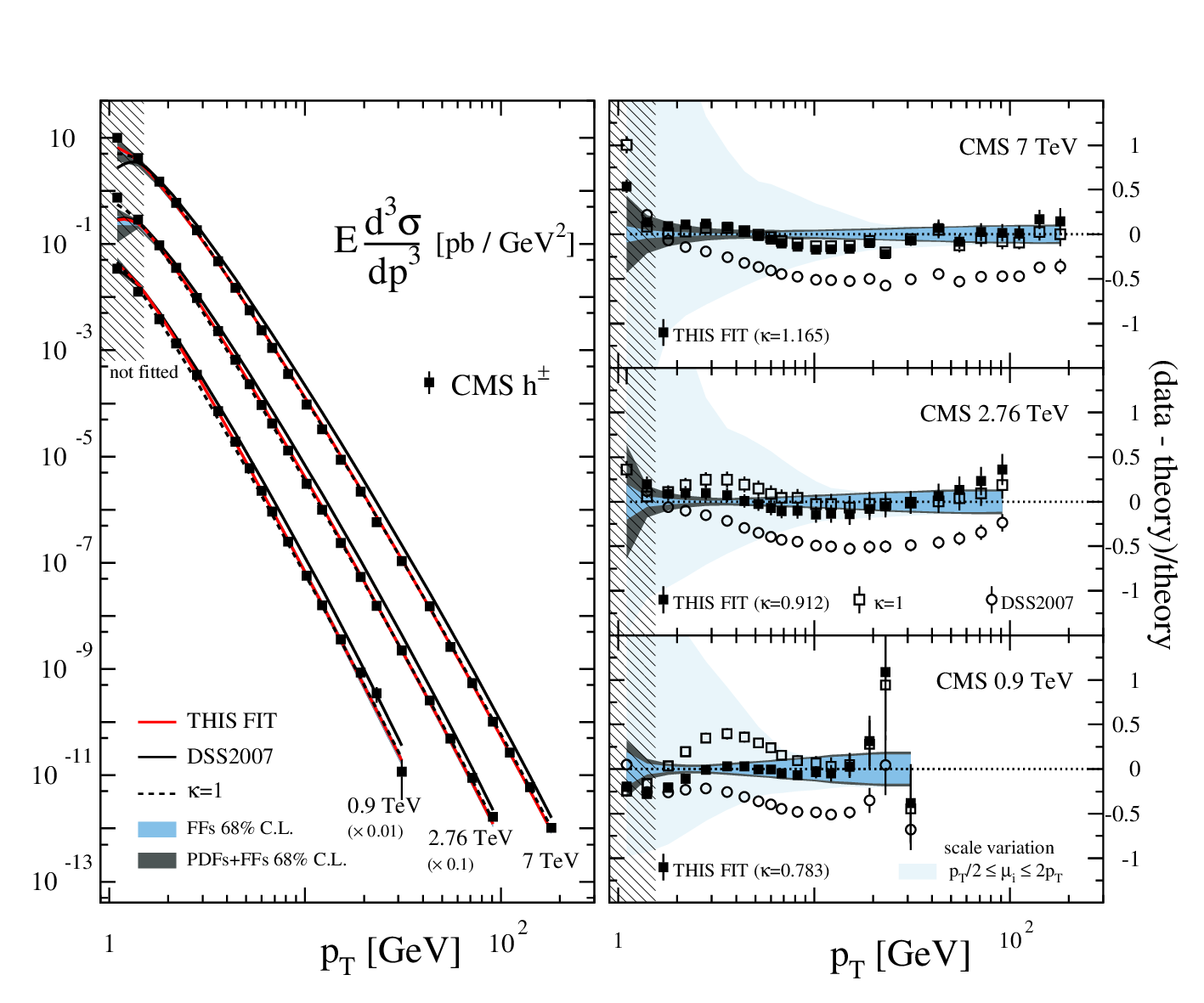,width=0.475\textwidth}
\vspace*{-0.4cm}
\caption{L.h.s: Comparison of our best fit and other NLO results with the CMS data on $h^{\pm}$ production \cite{CMS:2011mry,CMS:2012aa}
at different $\sqrt{s}$. 
R.h.s.: \textquotedblleft(data-theory)/theory" plots for each set of data. 
The relevant cuts and sources of uncertainties for our new fit are indicated by the shaded bands; see Sec.\ III below.
\label{fig:pp-h-cms}}
\end{figure}

\begin{figure}[h!]
\vspace*{-0.7cm}
\epsfig{figure=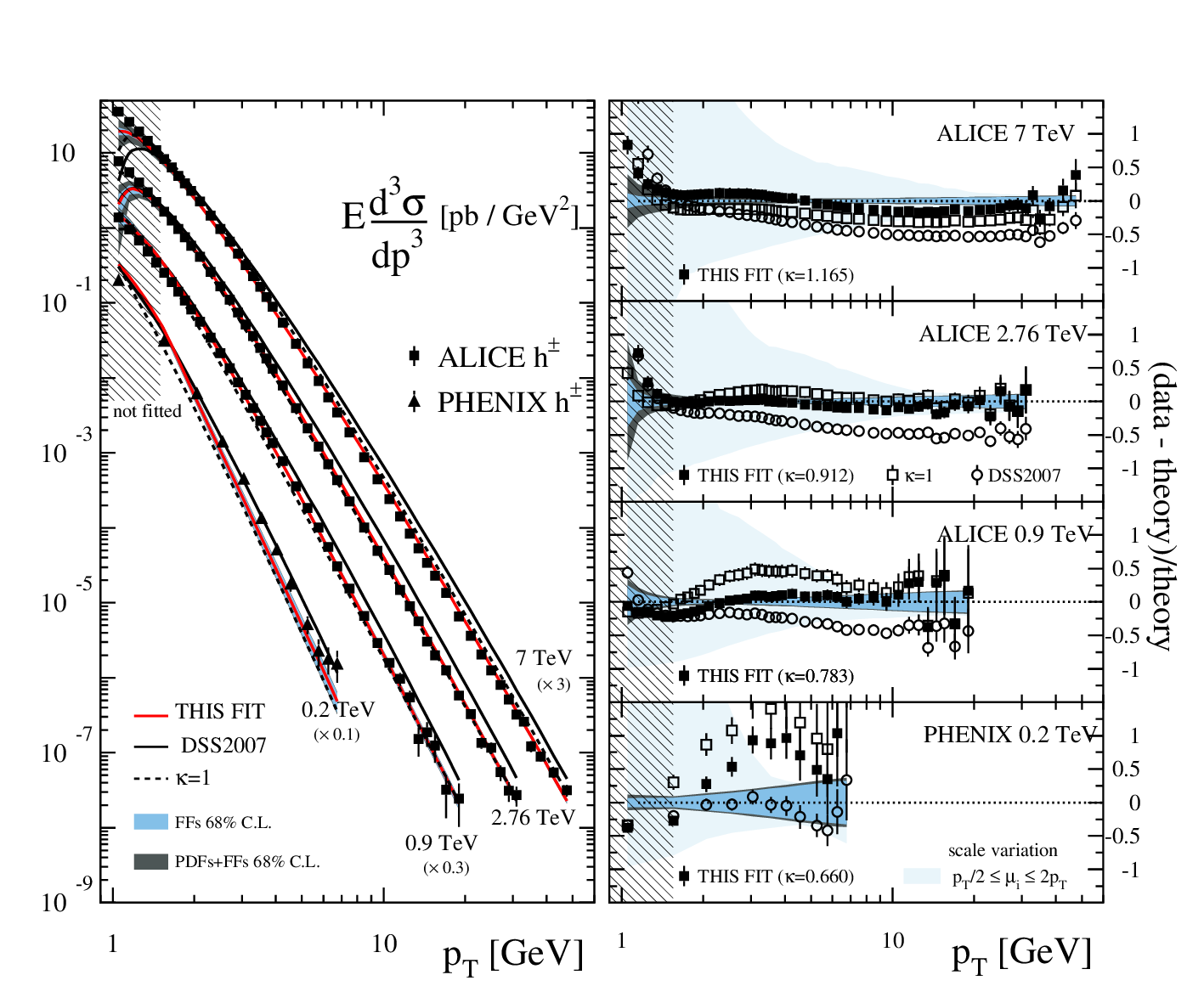,width=0.475\textwidth}
\vspace*{-0.4cm}
\caption{
As Fig.~\ref{fig:pp-h-cms} but now for both ALICE \cite{ALICE:2010mty,ALICE:2013txf} and PHENIX \cite{PHENIX:2005jxc} data.
\label{fig:pp-h-alice&phenix}}
\vspace*{-4mm}
\end{figure}

In Figures~\ref{fig:pp-h-cms} and \ref{fig:pp-h-alice&phenix} we compare CMS, ALICE,
and PHENIX \cite{PHENIX:2005jxc} data, all included in our global fit to be discussed below, 
against NLO estimates computed with parton distribution functions (PDFs) from
Ref.~\cite{Bailey:2020ooq} and different choices of FFs.
The estimates obtained with the pre-LHC set of FFs \cite{deFlorian:2007ekg} 
are represented by solid black lines and open circles
in the panels on left-hand-side (l.h.s.) and right-hand-side (r.h.s.) of both figures, respectively.
Clearly, these results labeled as DSS2007 overestimate the CMS and ALICE data in the entire range of $p_T$ 
shown in Figs.~\ref{fig:pp-h-cms} and \ref{fig:pp-h-alice&phenix} but, 
at the same time, describe the PHENIX data at $\sqrt{s}=200\,\mathrm{GeV}$,
included in their fit, very well.
%

\begin{figure}[t!]
\vspace*{-0.3cm}
\epsfig{figure=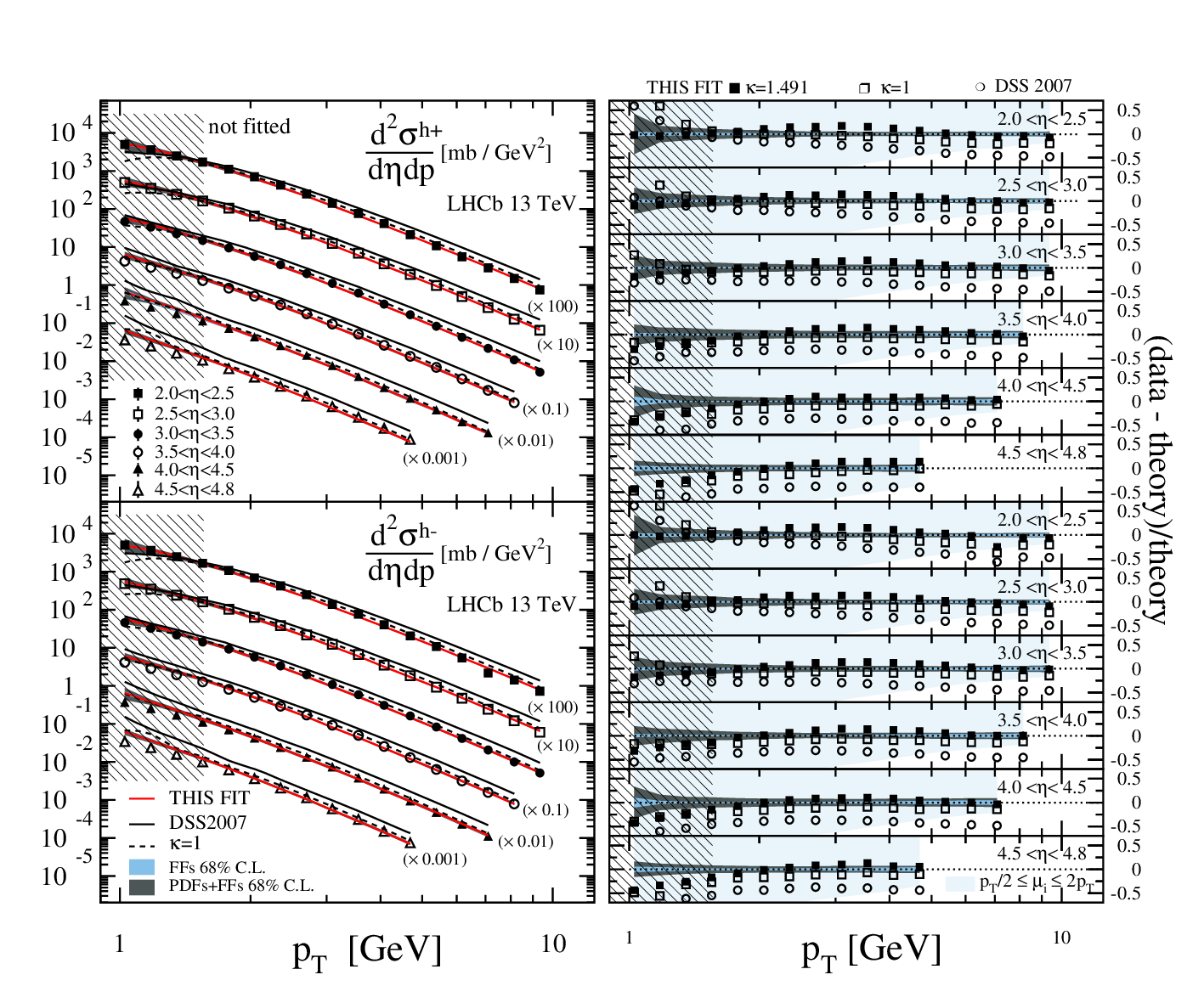,width=0.475\textwidth}
\vspace*{-0.4cm}
\caption{As Fig.~\ref{fig:pp-h-cms} but now for the charge separated $h^+$ and $h^-$ data from LHCb \cite{LHCb:2021abm} in
different bins of rapidity.
\label{fig:pp-h-lhcb}}
\end{figure}

In addition, the LHCb experiment has recently reported measurements at 
$\sqrt{s}=13\,\,\mbox{TeV}$ \cite{LHCb:2021abm} that discriminate positively and negatively charged hadrons $h^+$ and $h^-$
in various bins of (forward) rapidity $\eta$ up to almost 5 units. These results nicely complement 
the charged averaged $h^{\pm}$ ALICE and CMS data taken at central rapidities in our analysis.
Again, the NLO estimates computed with the DSS2007 FFs \cite{deFlorian:2007ekg} overestimate the data in the
entire ranges of $p_T$ and $\eta$ explored by LHCb as can be inferred from Fig.~\ref{fig:pp-h-lhcb}.
%
%
\begin{figure}[h!]
\vspace*{-0.7cm}
\epsfig{figure=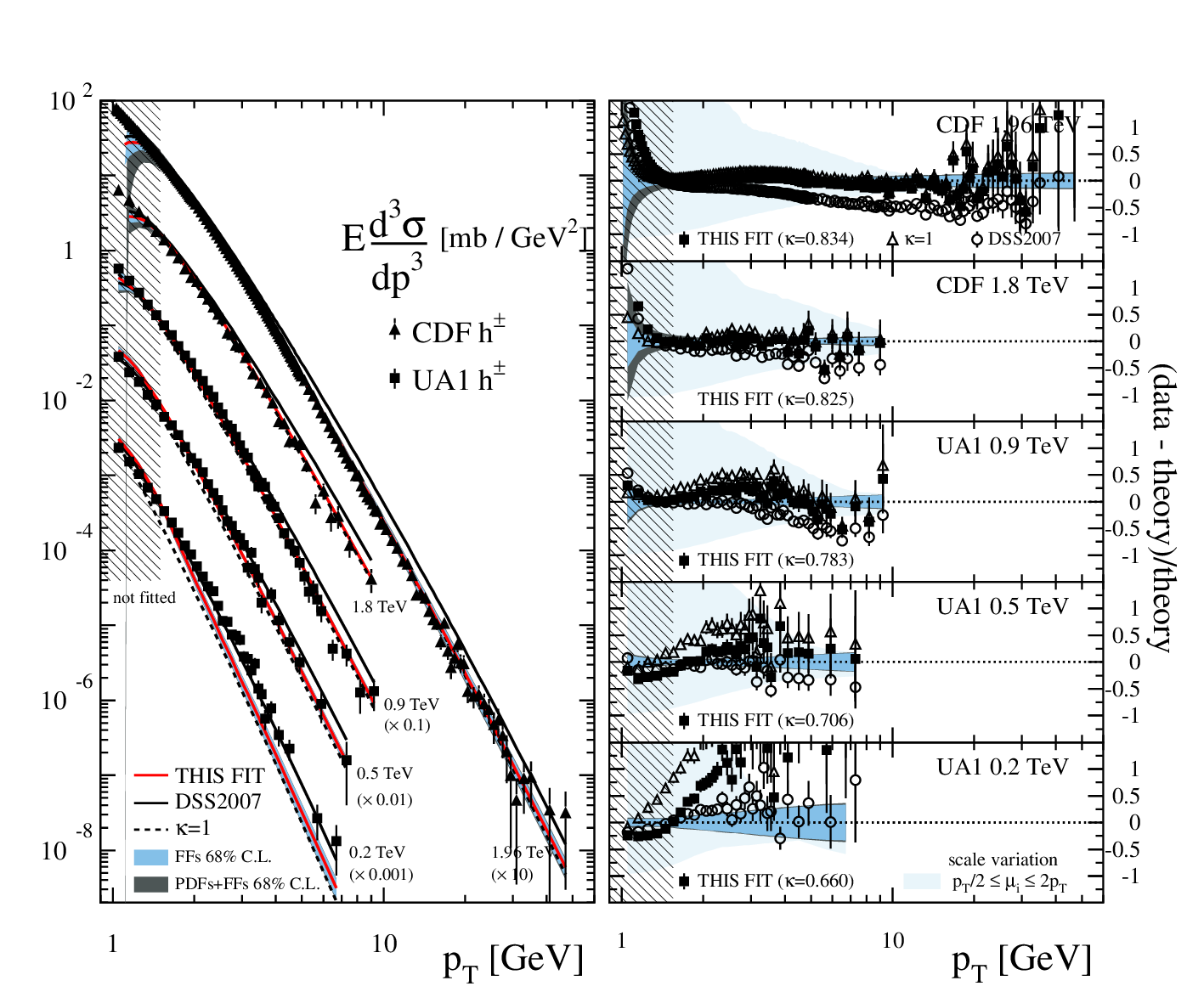,width=0.475\textwidth}
\vspace*{-0.4cm}
\caption{As Fig.~\ref{fig:pp-h-cms} but now for CDF \cite{CDF:1988evs,CDF:2009cxa} and UA1 \cite{UA1:1989bou} data.
\label{fig:pp-h-ua&cdf}}
\vspace*{-4mm}
\end{figure}

In our fit, we include also charge-averaged $h^{\pm}$ data taken in proton-antiproton collisions from both 
CDF and the UA1 and UA2 experiments, some of which are shown in Fig.~\ref{fig:pp-h-ua&cdf}. 
As before, the DSS2007 FFs tend to overestimate the data at larger $\sqrt{s}$, even though
some of them were included in their analysis \cite{deFlorian:2007ekg}. 
It is worth noticing that, at variance with the recent global analysis of pion FFs in Ref.~\cite{Borsa:2021ran}, 
where the only source of conflicting high energy collider data stems from a single LHC experiment (ALICE),
data sets from four experiments and from different colliders (LHC and Tevatron) show the same trend 
of NLO estimates grossly overshooting data in case of unidentified charged hadrons $h^{\pm}$.

\begin{figure}[t!]
\vspace*{-0.2cm}
\hspace*{-0.3cm}
\epsfig{figure=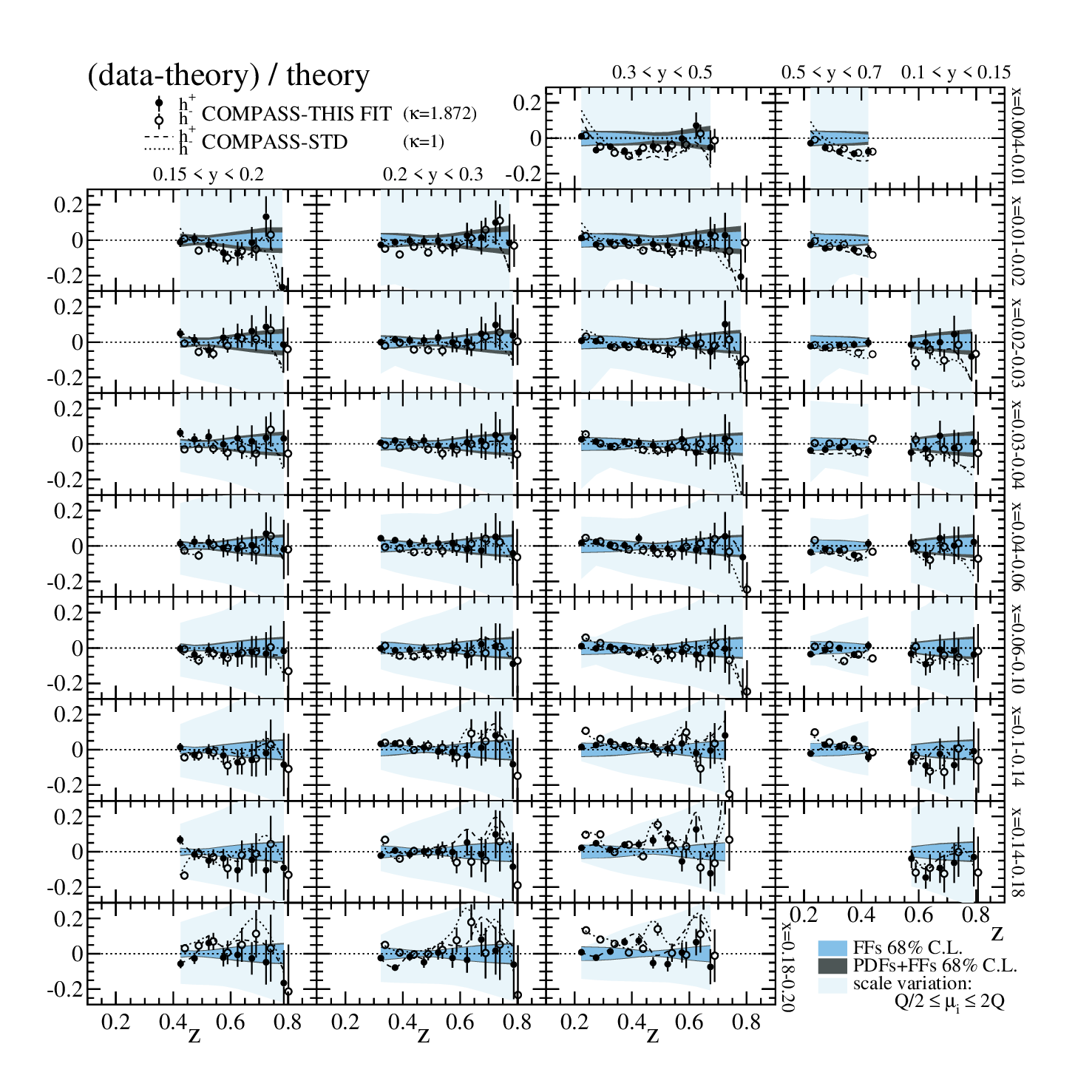,width=0.52\textwidth}
\vspace*{-0.4cm}
\caption{\textquotedblleft(data-theory)/theory" comparisons of the DSS2007 and our best fit results to
the charge separated $h^+$ and $h^-$ SIDIS multiplicities from COMPASS. The shaded bands provide
various uncertainty estimates, see Sec.\ III below.
\label{fig:compass}}
\vspace*{-4mm}
\end{figure}

Along with the high-$p_T$ hadroproduction results shown in Figs.~\ref{fig:pp-h-cms}-\ref{fig:pp-h-ua&cdf},
our new global QCD analysis of FFs for unidentified charged hadrons includes the same sets of
SIA data \cite{ref:tpcdata,ref:slddata,ref:alephdata,ref:delphidata,ref:opaleta,ref:tassodata,ref:opal,ref:opall,ref:delphil}
as in Ref.~\cite{deFlorian:2007ekg} as well as the SIDIS data on proton and deuteron targets from EMC \cite{EuropeanMuon:1991sne}.
A new, crucial asset in our fit are the much more precise SIDIS multiplicities for $h^+$ and $h^-$ production 
from COMPASS \cite{COMPASS:2016xvm}, which are presented in a very detailed binning in the relevant kinematical variables. 
These data are instrumental to achieve the charge and flavor separation of the $h^{\pm}$ FFs in our fit. 
In Fig.~\ref{fig:compass} we show a \textquotedblleft(data-theory)/theory" comparison of the charge separated SIDIS multiplicities 
from COMPASS for various theoretical estimates. Contrary to the PP data in Figs.~\ref{fig:pp-h-cms}-\ref{fig:pp-h-ua&cdf}, 
the DSS2007 FFs \cite{deFlorian:2007ekg} nicely reproduce the data. Again, the NLO estimates are computed with PDFs from Ref.~\cite{Bailey:2020ooq} neglecting 
nuclear effects in the deuteron \cite{Epele:1991np}.
%

The present analysis of the $h^{\pm}$ FFs differs in one important aspect from our previous
fit in Ref.~\cite{deFlorian:2007ekg}. In the latter case, the $h^{\pm}$ FFs were built up
from the sum of the previously determined charged pion, kaon, and (anti)proton FFs plus
a small residual component which was actually fitted to the available data.
Now we choose to parameterize the entire set of charged hadron FFs from scratch. 
The reasoning behind this is twofold. Firstly, the availability 
of many new and, most importantly, significantly more precise data allows us to
constrain and fully discriminate between the flavors of the parton-to-charged hadron FFs in the fit.
Secondly, it greatly facilitates the estimates of the remaining uncertainties of the obtained FFs
by avoiding a cumbersome propagation and interplay of uncertainties stemming from
pion, kaon, and (anti)proton FFs.
Nevertheless, the fits obtained in either way are equally good, and the residual contribution 
to the newly obtained $h^{\pm}$ FFs is indeed small and positive once the pion, kaon, and (anti)proton 
contributions are subtracted. In other words, the $h^{\pm}$ FFs are consistent with the individual contributions 
of each of the identified hadron species contributing to the sum of unidentified charged hadron yields.

Finally, following the strategy of Ref.~\cite{Borsa:2021ran} for the extraction of pion FFs, 
we perform two fits that differ in the choice for the 
renormalization, initial-, and final-state factorization scale,  
$\mu_{R}$, $\mu_{FI}$, and $\mu_{FF}$, respectively.
In one fit we adopt the standard choice, i.e., we set    
$\mu_{R}=\mu_{FI}=\mu_{FF}=\kappa\, {\cal E}$ with $\kappa=1$ and ${\cal E}$
being the typical hard scale of the process under consideration (e.g.\ $p_T$ in PP).
In our best fit, that gives a much better global description of the data sets entering the analysis,
we treat $\kappa$ as free parameter for each experiment or group of experiments with similar kinematics.
It turns out that the $\kappa$ values that have been obtained in the analysis of pion FFs \cite{Borsa:2021ran}
also lead to a very close to optimum description of the charged hadron data. 
Therefore, we simply adopt the $\kappa$ values from the pion fit rather than refitting them.
A similar observation can be made for fits of charged kaon and (anti)proton FFs which will be
reported elsewhere. 
We also replicate the strategy of Ref.~\cite{Borsa:2021ran} for the estimation of the uncertainties of the resulting set of FFs. 
Using the optimum values of the fit parameters $\kappa$, and through the application of a Monte Carlo sampling technique, we generate 
an ensemble of $N_{\mathrm{rep}}=500$ replicas of the FFs. The uncertainty stemming from the FFs in any observable is then assumed to be given 
statistically by the standard deviation of the observable calculated over the set of replicas. 
%
\section{Discussion of the Fit Results}
The hadroproduction cross section estimates computed with our new optimum set of
charged hadron FFs, i.e.\ obtained from the fit by adopting the scale factors $\kappa$ 
from the pion fit \cite{Borsa:2021ran}, are shown as red solid lines
on the l.h.s.\ of Figs.~\ref{fig:pp-h-cms}-\ref{fig:pp-h-ua&cdf}, 
and by filled rectangles in the corresponding \textquotedblleft(data-theory)/theory" plots
on the r.h.s.\ of each figure.
The alternative results calculated with the standard choice $\kappa=1$ are represented 
by dashed lines and open rectangles, respectively. 
In addition to the NLO estimates, Figs.~\ref{fig:pp-h-cms}-\ref{fig:pp-h-ua&cdf} also present the uncertainty bands associated 
with the nonperturbative inputs of the calculations at the $68\%$ confidence level (C.L.), 
i.e.\ from the used sets of FFs and PDFs (blue and grey shaded bands), as well as the theoretical uncertainty stemming from the truncation 
of the perturbative series (light blue shaded bands). 
The latter is estimated performing the so-called 27-point independent variation of the renormalization and factorization scales 
in the range $p_T/2\leq\mu\leq2 p_T$.

In contrast to the analysis of Ref.~\cite{deFlorian:2007ekg}, the fit to the PP data set is now primarily driven by the very precise measurements 
taken at higher c.m.s.\ energies of the LHC. As can be seen, the $\kappa=1$ fit reproduces fairly well the data at energies of around 
$\sqrt{s} = 7\,\text{and}\,13$ TeV, while the quality of the fit degrades sharply towards smaller values of $\sqrt{s}$. 
The use of the scale variable $\kappa$, on the other hand, mitigates some of the tension leading to a nice description of 
hadroproduction data from the highest c.m.s.\ energies down to $\sqrt{s}\simeq 0.5$ TeV, and significantly improves the agreement 
with data at $\sqrt{s} \simeq 0.2$ TeV as compared to the $\kappa=1$ results. In the latter case, the DSS2007 fit still provides 
the best results but, as previously mentioned, grossly overestimates all LHC data. 

We note that for SIA and SIDIS the agreement of data with the theoretical estimates 
is to a much lesser extent affected by the choice of scales than calculations for PP.
In the SIDIS multiplicities the dependence on the scales $\mu_{R}$, $\mu_{FI}$, and $\mu_{FF}$
tends to cancel in the relevant ratios of the semi-inclusive and the inclusive cross section. 
For completeness, the optimum values of $\kappa$ for each set of data are indicated in 
Figs.~\ref{fig:pp-h-cms}-\ref{fig:compass}.

\begin{figure}[t!]
\vspace*{-0.5cm}
\hspace*{-0.3cm} 
\epsfig{figure=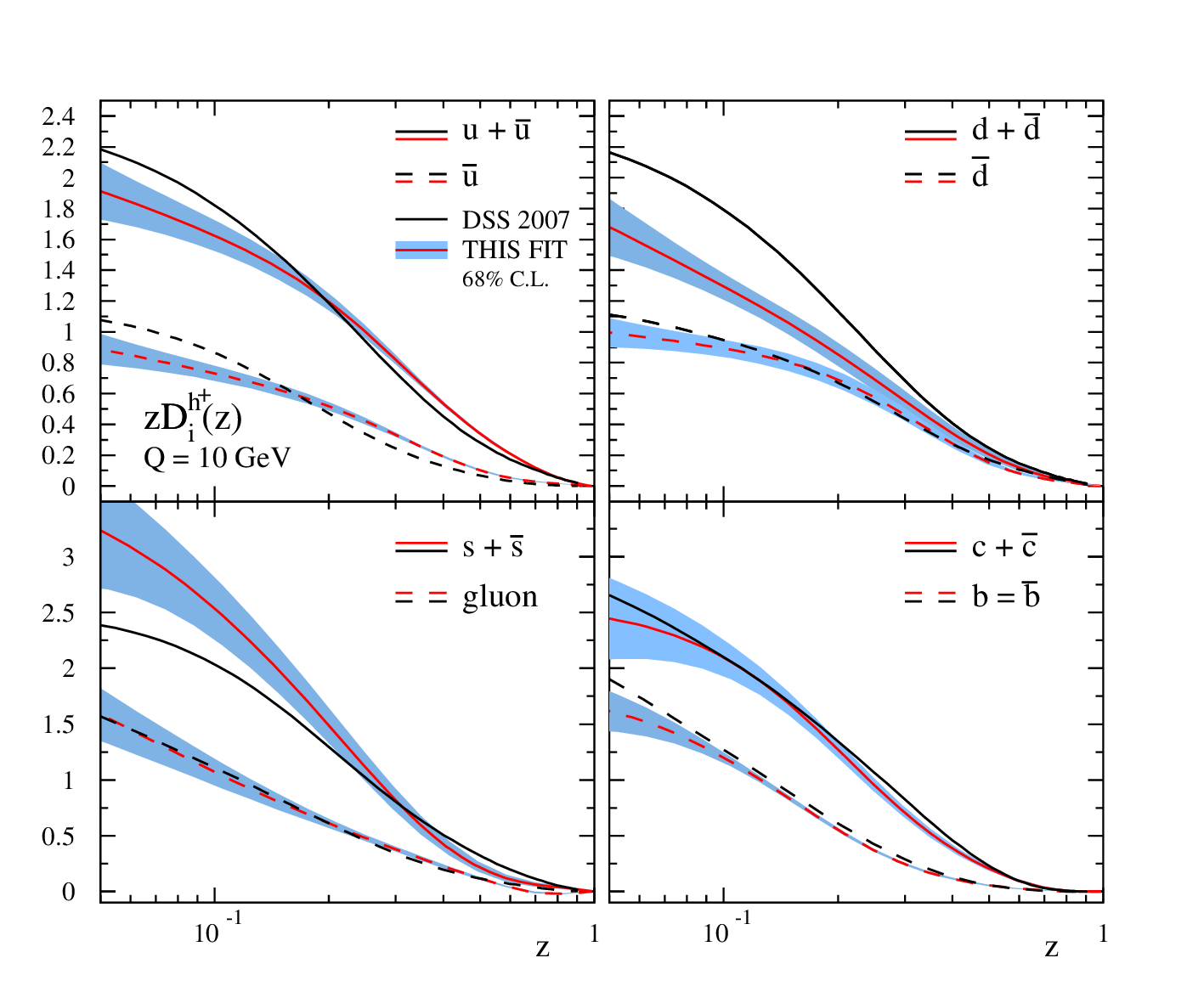,width=0.50\textwidth}
\vspace*{-0.8cm}
\caption{The obtained optimum FFs (red lines) for positively charged hadrons
for all parton flavors at scale $Q=10\,\mathrm{GeV}$ along with
uncertainty estimates at 68\% C.L.\ and compared to the previous results (black lines)
from DSS2007 \cite{deFlorian:2007ekg}.
\label{fig:dist}}
\vspace*{-3mm}
\end{figure}
In Fig.~\ref{fig:dist} we compare the newly obtained optimum set of FFs
$zD_i^{h^{+}}(z,Q^2)$ for positively charged hadrons, multiplied by the momentum fraction $z$,
at a scale $Q=10\,\mathrm{GeV}$ for all parton flavors $i$
to the previous analysis by DSS .
The shaded band indicate our estimates for the uncertainties of
the FFs for each flavor at the 68\% confidence level.

As can be seen, there are sizable differences between the current and the DSS results \cite{deFlorian:2007ekg}
for all light quark flavor combinations $q+\bar{q}$, where $q=u,\,d,\,s$. 
In case of $u+\bar{u}$, they originate largely from the unfavored flavor $\bar{u}$,
where favored and unfavored refer to the valence content of a $\pi^+$ meson.
The fragmentation into a $\pi^+$ contributes by far the most to the sum 
of FFs into a positively charged hadron $h^+$.
Likewise, we find that for $d+\bar{d}$, the differences for the favored
contribution from $\bar{d}$ are much more moderate than for the unfavored $d$.
We recall that the flavor separation in the DSS2007 extraction \cite{deFlorian:2007ekg}
was mainly driven by the EMC SIDIS data \cite{EuropeanMuon:1991sne}, and, to some extent, 
by preliminary Hermes results used to constrain pion and kaon FFs \cite{deFlorian:2007aj}. 
In the current fit, the separation of the $h^{\pm}$ FFs into different parton flavors
comes mostly from the much more precise COMPASS data \cite{COMPASS:2016xvm} shown in Fig.~\ref{fig:compass}. 
It should be kept in mind though, that SIDIS data are only available for hadron momentum fractions 
$z\ge 0.2$. Hence, for smaller values of $z$, the flavor separation is at best an 
extrapolation driven by constraints at higher $z$. 

On the other hand, the best experimental information on heavy flavor FFs still
stems from flavor-tagged SIA data. So it is not surprising, that the differences
for charm and bottom FFs, shown in the lower right panel of Fig.~\ref{fig:dist},
are very small. What comes as a surprise, is the similarity between the 
gluon FF in both fits. In the DSS2007 analysis the gluon FF followed from a compromise 
between UA1/UA2 and the Tevatron data, while the current fit makes use of a
wealth of LHC hadroproduction data. We therefore conclude, that the observed
similarity is most likely purely accidental.

\begin{table}[t!]
\caption{\label{tab:exppiontab} Data sets, normalizations $N_i$ as defined in Eq.~(6) of \cite{deFlorian:2014xna}, 
and the partial and total $\chi^2$ values obtained in the fit.}
\begin{ruledtabular}
\begin{tabular}{lcccc}
experiment& data & rel.\ norm.\ &data  & $\chi^2$ \\
          & type & in fit       & fitted     &         \\\hline
TPC \cite{ref:tpcdata}  & incl.\  &  0.983 & 17 & 11.6 \\    
SLD \cite{ref:slddata}  & incl.\  &  1.026 & 21 & 18.0 \\
ALEPH \cite{ref:alephdata}& incl.\ & 1.028 & 27 & 19.6 \\
DELPHI \cite{ref:delphidata}  & incl.\  & 1.048  & 12 & 8.3 \\
          & ``$uds$ tag''   &  1.048  & 12 & 15.0 \\
          & ``$b$ tag''     &  1.048  & 12 & 2.1  \\
TASSO \cite{ref:tassodata}  & incl.\ (44 GeV)  & 1.035  & 14 & 10.5 \\
                            & incl.\ (35 GeV)  & 1.035  & 14 & 15.9 \\
OPAL \cite{ref:opal}   & incl.\  & 1.057  & 12 & 6.5 \\
          & ``$uds$ tag'' &  1.057  & 12 & 10.4 \\
          & ``$c$ tag'' &  1.057  & 12 & 6.4  \\
          & ``$b$ tag'' &  1.057  & 12 & 3.5  \\ 
ALEPH \cite{ref:alephdata}&  incl.\ long.\ & 1.028 & 11 & 2.5 \\
OPAL  \cite{ref:opall}   &  incl.\ long.\  & 1.057  & 12 & 3.0  \\
DELPHI \cite{ref:delphil}  & incl.\ long.\  & 1.048  & 12 & 13.2 \\
                          &  ``$uds$ tag long.'' & 1.048  & 12 & 38.6\\
                           &  ``$b$ tag long.'' &  1.048  & 12 & 5.0 \\ \hline
{\bf SIA data  }            &                    &       & 236   & 190.1   \\ \hline
EMC \cite{EuropeanMuon:1991sne}  & $p-h^+$    &  1.28 & 108 & 102.9 \\
                                 & $p-h^-$    &  1.28 & 108 & 223.3 \\
                                 & $d-h^+$    &  1.45 & 116 & 118.3 \\
                                 & $d-h^-$    &  1.45 & 116 & 239.9 \\

COMPASS \cite{COMPASS:2016xvm}  & $d-h^+$    &  0.989 & 311 & 179.4 \\
                                     & $d-h^-$ &  0.989 & 311  & 129.1 \\ \hline
 {\bf SIDIS data }            &                    &    & 1070 & 992.9   \\\hline
PHENIX \cite{PHENIX:2005jxc} &   0.20 TeV & 1.066 & 11 &  90.2      \\ 
UA1 \cite{UA1:1989bou}       &   0.20 TeV & 1.451 & 27 &  121.6     \\
                             &   0.50 TeV & 1.068 & 27 &  30.5      \\
                             &   0.63 TeV & 1.592 & 38 &  142.5     \\
                             &   0.90 TeV & 1.094 & 35 &  45.5      \\
UA2 \cite{UA2:1984ida}       &   0.54 TeV & 0.975 & 27 &  70.9      \\
CDF  \cite{CDF:1988evs}      &   0.63 TeV & 0.777 & 38 &  15.1      \\
\,\,\,\,\,\,\,\,\,\,\,\,\,   \cite{CDF:1988evs}    &   1.80 TeV & 1.974 & 32 &  41.2      \\
\,\,\,\,\,\,\,\,\,\,\,\,\, \cite{CDF:2009cxa}   & 1.96 TeV & 1.010 & 145 & 219.1      \\        
CMS \cite{CMS:2011mry}&    0.90 TeV & 0.947 & 15 &  38.7              \\
\,\,\,\,\,\,\,\,\,\,\,\,\,    \cite{CMS:2012aa} &    2.76 TeV & 0.999 & 19 &  23.6              \\
\,\,\,\,\,\,\,\,\,\,\,\,\,    \cite{CMS:2011mry}&    7.00 TeV & 0.948 & 22 &  91.7               \\
ALICE \cite{ALICE:2013txf} &    0.90 TeV & 0.981 & 32 &  69.3              \\
\,\,\,\,\,\,\,\,\,\,\,\,\,\,\,\,\,\,      \cite{ALICE:2013txf} &    2.76 TeV & 0.954 & 37 &  23.1               \\
\,\,\,\,\,\,\,\,\,\,\,\,\,\,\,\,\,\,     \cite{ALICE:2010mty} &    7.00 TeV & 0.948 & 43 &  92.3               \\
LHCb \cite{LHCb:2021abm}  &    13.0 TeV & 1.034 & 140 & 775.5               \\ \hline
 {\bf PP data }            &                    &       & 688  & 1890.8   \\ 
 \hline\hline
{\bf TOTAL:} & & & 1994  & 3073.8 \\
\end{tabular}
\end{ruledtabular}
\end{table}
Finally, in Table~\ref{tab:exppiontab} we summarize the data sets used in the current 
NLO global QCD analysis, the computed normalization shifts $N_i$ as defined in Eq.~(6) of Ref.~\cite{deFlorian:2014xna}, 
and the corresponding partial $\chi^2$-values. 
As in all previous fits, \cite{deFlorian:2007ekg,Borsa:2021ran,deFlorian:2014xna}, we include only SIA data
for hadron momentum fractions $z>0.1$ since the hadrons are treated as massless in the factorized QCD framework.
All the contributing data sets from SIA are very well reproduced with the possible exception of the light quark flavor-tagged 
data from DELPHI.
As was mentioned above, SIDIS data are only available for $z>0.2$. It turns out that the very precise, new COMPASS data 
are remarkably well reproduced by the fit while the old EMC data for negatively charged hadrons have a rather large
partial $\chi^2$ and also acquire large normalization shifts. However, most of the large $\chi^2$ contribution
stems from just a few data points.
Hadroproduction data taken in proton-(anti)proton collisions give the largest contribution to the total $\chi^2$ of the fit.
Here we adopt a cut $p_T>1.5\,\mathrm{GeV}$ in the transverse momentum of the detected charged hadron for all data sets except for
LHCb, where we require $p_T > 1.56\,\mathrm{GeV}$ to limit unwanted uncertainties from the PDFs, propagating into 
the cross section calculations, to a reasonable level. 
While the overall description of the PP data is good and much improved as compared to calculations based on the
DSS2007 set of charged hadron FFs, it is certainly not perfect. Some older sets from UA1 and CDF receive large normalization shifts,
which is most likely an indication that these data lead to unresolvable tensions when combined with other measurements;
similarly for the EMC SIDIS data for negatively charged hadrons.
The LHC data, which span a large energy and $p_T$-range and, in case of LHCb, also a fairly extreme kinematical coverage in terms of rapidity,
are all remarkably well reproduced in the fit along with other probes from SIA and SIDIS. It should be also kept in mind, that
the scale uncertainties for the PP cross section estimates at NLO accuracy are extremely large below $p_T\simeq 5\,\mathrm{GeV}$ 
as can be seen in the panels on r.h.s. of Figs.~\ref{fig:pp-h-cms}-\ref{fig:pp-h-lhcb}. 

\section{Summary and Conclusions}
We have presented an updated set of parton-to-unidentified charged hadron fragmentation functions at NLO accuracy.
It reproduces  the latest LHC hadroproduction data at different c.m.s.\ energies and down
to transverse momentum values of about $1.5\,\mathrm{GeV}$ in a global analysis
together with results from SIA, SIDIS, and older hadroproduction data taken in proton-(anti)proton collisions
at various energies.
As in the recent analysis of pion FFs in Ref.~\cite{Borsa:2021ran}, the best fit
exploits the freedom in the choice of the renormalization and factorization scales.  
In addition, a second set of FFs with conventional factorization and renormalization scale choices is provided.

The new analysis supersedes the extraction of FFs presented in Ref.~\cite{deFlorian:2007ekg} in many ways.
Firstly, because of the inclusion of the latest experimental information from the LHC and the COMPASS SIDIS data,
and, secondly, in the way uncertainties are estimated and presented through Monte Carlo replicas.
The latter allow one to easily propagate the obtained uncertainties of the charged hadron FFs to any observable 
depending on them. 

The significance of the factorization scale dependence and the need to utilize it in the fit
points to a limitation of the NLO approximation, which is much more pronounced in proton-(anti)proton collisions 
than in the other processes studied. 
Interestingly and reassuringly, the scale choices that optimize the unidentified charged hadron FFs in our fit
are indistinguishable from those found for pions in another recent global analysis. 
This is to be expected as charged hadrons produced in hard proton-(anti)proton collisions are dominated by pions.

We believe that the current NLO analysis provides a useful and up-to-date tool for phenomenological studies 
involving charged hadron FFs. These include not only hadroproduction in proton-proton and proton-nuclei collisions, but also SIDIS measurements at the future Electron Ion Collider \cite{Aschenauer:2019kzf} and as a way to further constrain PDFs at the
comparatively lower energy scales typical of SIDIS experiments \cite{Borsa:2017vwy}. 
Despite its shortcomings in describing the energy dependence of 
currently available PP data in all its details, it accurately reproduces the main features 
of the different probes adopted in the global analysis. 
A better understanding of single-inclusive hadroproduction cross sections
can be only expected once the full QCD corrections at next-to-next-to-leading order accuracy
become available. 
Then it remains to be seen whether the theoretical description and the scale dependence
will improve or if the underlying framework has some serious shortcomings such as significant
factorization breaking effects.

%
This work was supported in part by CONICET, ANPCyT, UBACyT, 
the Bundesministerium für Bildung und Forschung (BMBF), grant 05P21VTCAA,
and the Deutsche Forschungsgemeinschaft (DFG) 
through the Research Unit FOR 2926 (Project No.\ 40824754). 
%

\end{document}